\documentclass[twocolumn,aps,prl,showpacs]{revtex4}

\begin{document}
\title{Stronger Two-Observer All-Versus-Nothing Violation of Local Realism}
\author{Ad\'{a}n Cabello}
\email{adan@us.es}
\affiliation{Departamento de F\'{\i}sica Aplicada II,
Universidad de Sevilla, 41012 Sevilla, Spain}
\date{\today}


\begin{abstract}
We introduce a two-observer all-versus-nothing proof of Bell's
theorem which reduces the number of required quantum predictions
from 9 [A. Cabello, Phys. Rev. Lett. {\bf 87}, 010403 (2001);
Z.-B. Chen {\em et al.}, Phys. Rev. Lett. {\bf 90}, 160408 (2003)]
to 4, provides a greater amount of evidence against local realism,
reduces the detection efficiency requirements for a conclusive
experimental test of Bell's theorem, and leads to a Bell's
inequality which resembles Mermin's inequality for three observers
[N. D. Mermin, Phys. Rev. Lett. {\bf 65}, 1838 (1990)] but
requires only two observers.
\end{abstract}


\maketitle


The Greenberger-Horne-Zeilinger (GHZ) proof~\cite{GHZ89,Mermin90a}
of Bell's theorem~\cite{Bell64} provided a direct contradiction
between the Einstein-Podolsky-Rosen (EPR)~\cite{EPR35} local
elements of reality (LERs) by considering 4~perfect correlations
predicted by quantum mechanics. However, while Bell's theorem
required only two spacelike separated observers, the GHZ proof
required three. A two-observer all-versus-nothing (AVN)
proof~\cite{Mermin90b} based on 9 perfect correlations was
introduced in Refs.~\cite{Cabello01a,Cabello01b}, then adapted for
two-photon systems~\cite{CPZBZ03}, and recently tested in the
laboratory~\cite{YZZYZZCP05}. Despite requiring only two
observers, this AVN proof has two disadvantages when compared to
GHZ: it is more complex, in the sense that it requires a higher
number of quantum predictions, and provides less evidence against
local realism than GHZ~\cite{VGG03}; thus a conclusive experiment
based on the two-observer proof would require higher detection
efficiencies than one based on GHZ. Both disadvantages are indeed
related: in the two-observer proof, 8 out of these 9 perfect
correlations (i.e., 89\%) can be explained by LERs; in GHZ, only 3
out of these 4 perfect correlations (i.e., only 75\%) can be
explained by LERs.

Here we introduce an AVN proof requiring two observers and only 4
perfect correlations. We then show that conclusive experiments
based on this new proof would require lower detection efficiencies
than those needed for the previous two-observer AVN proof, and
derive the corresponding Bell inequality for real experiments.

Let us consider a two-photon system entangled both in polarization
and path degrees of
freedom~\cite{CPZBZ03,YZZYZZCP05,Kwiat97} prepared in the
state
\begin{eqnarray}
|\psi\rangle & = & \frac{1}{2} ( |H u\rangle_1 |H u\rangle_2 +
|H d\rangle_1 |H d\rangle_2 \nonumber \\
& & + |V u\rangle_1 |V u\rangle_2 - |V d\rangle_1 |V d\rangle_2),
\label{benasque}
\end{eqnarray}
where $|H\rangle_j$ and $|V\rangle_j$ represent horizontal and
vertical polarization, and $|u\rangle_j$ and $|d\rangle_j$ denote
two orthonormal path states for photon~$j$. The
state~(\ref{benasque}) can be viewed as a two-photon version of
the four-qubit cluster state~\cite{BR01}. Lu~\cite{Lu05a} and
Lu~\cite{Lu05b} have described procedures for preparing
state~(\ref{benasque}).

Let us also consider 6~local observables on photon~$j$: 3 for
polarization degrees of freedom, defined by the operators
\begin{eqnarray}
X_j & = & |H\rangle_j \langle V|+|V\rangle_j \langle H|, \\
Y_j & = & i \left(|V\rangle_j \langle H|-|H\rangle_j \langle V|\right), \\
Z_j & = & |H\rangle_j \langle H|-|V\rangle_j \langle V|,
\end{eqnarray}
and 3 for path degrees of freedom, defined by the operators
\begin{eqnarray}
x_j & = & |u\rangle_j \langle d|+|d\rangle_j \langle u|, \\
y_j & = & i \left(|d\rangle_j \langle u|-|u\rangle_j \langle
d|\right),
\\
z_j & = & |u\rangle_j \langle u|-|d\rangle_j \langle d|.
\end{eqnarray}
Each of these observables can have only one of two possible
values: $-1$ or $1$.

The proof has two steps. First, we will show that the 7~local
observables $X_1$, $Y_1$, $x_1$, $X_2$, $Y_2$, $y_2$, and $z_2$
satisfy the EPR condition for LER; namely, ``if, without in any
way disturbing a system, we can predict with certainty (i.e., with
probability equal to unity) the value of a physical quantity, then
there exists an element of physical reality corresponding to this
physical quantity''~\cite{EPR35}. The observables $X_1$, $Y_1$,
and $x_1$ of photon~1 are LERs because their values can be
predicted with certainty from spacelike separated measurements on
photon~2. Spacelike separation guarantees that photon~1 has not
been disturbed in any way~\cite{Einstein48}. These predictions
with certainty follow from the fact that state~(\ref{benasque})
satisfies the following equations:
\begin{eqnarray}
X_1 X_2 z_2 |\psi \rangle & = & |\psi \rangle,
\label{equno} \\
Y_1 Y_2 z_2 |\psi \rangle & = & -|\psi \rangle,
\label{eqcinco} \\
x_1 Z_2 x_2 |\psi \rangle & = & |\psi \rangle.
\label{eqdos}
\end{eqnarray}
Equation~(\ref{equno}) tells us that, from the results of
measuring $X_2$ and $z_2$ on photon~2, we can predict with
certainty the result $v(X_1)$ of measuring $X_1$ on photon~1.
Equation~(\ref{eqcinco}) tells us that, from the results of
measuring $Y_2$ and $z_2$ on photon~2, we can predict with
certainty the result $v(Y_1)$ of measuring $Y_1$ on photon~1.
Equation~(\ref{eqdos}) tells us that, from the results of
measuring $Z_2$ and $x_2$ on photon~2, we can predict with
certainty the result $v(x_1)$ of measuring $x_1$ on photon~1.

Analogously, the observables $X_2$, $Y_2$, $y_2$, and $z_2$ of
photon~2 are also LERs because state~(\ref{benasque}) also
satisfies the following equations:
\begin{eqnarray}
X_1 z_1 X_2 |\psi \rangle & = & |\psi \rangle,
\label{eqtres} \\
Y_1 z_1 Y_2 |\psi \rangle & = & -|\psi \rangle,
\label{eqsiete} \\
Z_1 y_1 y_2 |\psi \rangle & = & -|\psi \rangle,
\label{eqocho} \\
z_1 z_2 |\psi \rangle & = & |\psi \rangle.
\label{eqmenosuno}
\end{eqnarray}
Therefore, the results $v(X_2)$, $v(Y_2)$, $v(y_2)$, and $v(z_2)$
of measuring $X_2$, $Y_2$, $y_2$, and $z_2$ on photon~2 can be
predicted with certainty from spacelike separated measurements on
photon~1.

Moreover, we can prove that two observables on the same photon,
but corresponding to different degrees of freedom, like $X_2$ and
$z_2$, are {\em independent} LERs in the sense that measuring one
of them does not change the value of the other (thus there is no
need for any additional assumptions beyond the EPR condition;
see~\cite{Marinatto03} for a similar discussion). For instance, a
suitable measurement of $X_2$ does not change $v(z_2)$ because
$v(z_2)$ can be predicted with certainty from a spacelike
separated measurement on photon~1 [see Eq.~(\ref{eqmenosuno})],
and this prediction is not affected by whether $X_2$ is measured
before measuring $z_2$, or $X_2$ and $z_2$ are jointly measured.
Therefore, the EPR criterion is enough to guarantee that $z_2$ has
a value $v(z_2)$, which does not change by measuring $X_2$. A
similar reasoning applies to all 7~local observables involved in
the proof.

The second step of the proof consists of showing the contradiction
between the predictions of quantum mechanics and those of local
realistic theories. For this purpose, note that
state~(\ref{benasque}) also satisfies the following equations:
\begin{eqnarray}
X_1 x_1 Y_2 y_2 |\psi \rangle & = & |\psi \rangle,
\label{eqnueve} \\
Y_1 x_1 X_2 y_2 |\psi \rangle & = & |\psi \rangle.
\label{eqonce}
\end{eqnarray}
To be consistent with Eqs.~(\ref{equno}), (\ref{eqcinco}),
(\ref{eqnueve}), and (\ref{eqonce}), local realistic theories
predict the following relations between the values of the LERs:
\begin{eqnarray}
v(X_1) & = & v(X_2) v(z_2),
\label{valuno} \\
v(Y_1) & = & -v(Y_2) v(z_2),
\label{valcinco} \\
v(X_1) v(x_1) & = & v(Y_2) v(y_2),
\label{valnueve} \\
v(Y_1) v(x_1) & = & v(X_2) v(y_2),
\label{valonce}
\end{eqnarray}
respectively. However, it is impossible to assign the values $-1$
or $1$ in a way consistent with all
Eqs.~(\ref{valuno})--(\ref{valonce}). This can be proved as
follows: In Eqs.~(\ref{valuno})--(\ref{valonce}) every value
appears twice; therefore, the product of
Eqs.~(\ref{valuno})--(\ref{valonce}) gives $1=-1$. We therefore
conclude that the 4~predictions of quantum mechanics given by
Eqs.~(\ref{equno}), (\ref{eqcinco}), (\ref{eqnueve}), and
(\ref{eqonce}) cannot be reproduced by EPR LERs.

The remarkable property of this AVN proof is that the
contradiction appears after considering only 4~quantum
predictions, while the previous two-observer AVN
proof~\cite{Cabello01a,Cabello01b,CPZBZ03,YZZYZZCP05} required
9~quantum predictions. Moreover, while a local realistic model can
reproduce 8 out of the 9~predictions of the previous two-observer
AVN proof, it can be easily seen that it can reproduce only 3 out
of the 4~predictions in the proof presented here. In more
practical terms, this means that an experimental realization of
the previous AVN proof~\cite{YZZYZZCP05} requires photodetectors
of a higher efficiency to avoid the detection
loophole~\cite{Pearle70} than an experimental realization of the
proposed proof. To show this, we will estimate the detector
efficiency required for both proofs. For this purpose, it is
useful to see both proofs as games~\cite{Vaidman99}.

Let us start by translating the new proof into a game in which a
quantum-based strategy beats any classical strategy. Consider a
team of two players, Alice and Bob, each of them isolated in a
booth. Alice is asked one out of two possible questions: (I)
``What are $v(X_1)$ and $v(x_1)$?'' or (II) ``What are $v(Y_1)$
and $v(x_1)$?'' Bob is asked one out of 4 possible questions: (i)
``What are $v(X_2)$ and $v(y_2)$?,'' (ii) ``What are $v(X_2)$ and
$v(z_2)$?,'' (iii) ``What are $v(Y_2)$ and $v(y_2)$?,'' or (iv)
``What are $v(Y_2)$ and $v(z_2)$?'' Each of them must give one of
the following answers: ``$-1$ and $-1$,'' ``$-1$ and $1$,'' ``$1$
and $-1$,'' or ``$1$ and $1$.'' When Alice is asked (I), Bob is
asked (ii) or (iii); when Alice is asked (II), Bob is asked (i) or
(iv). Since $v(X_2)$ represents a LER, Bob's answer to $v(X_2)$
must be independent on whether $v(X_2)$ is asked together with
$v(y_2)$ or with $v(z_2)$. The same applies for Bob's answers to
$v(Y_2)$, $v(y_2)$, and $v(z_2)$. The team wins if their answers
satisfy the corresponding equation of the set
(\ref{valuno})--(\ref{valonce}).

Assuming that the 4~possible combinations of questions are asked
with the same frequency, no classical strategy allows the players
to win in more than~$3/4$ of the rounds. For instance, a simple
optimal classical strategy is that each player always answers~$1$
to any question. The hidden set of local instructions
\begin{eqnarray}
\left\{ \begin{array}{c|cc}
v(X_1) & v(X_2) & v(z_2) \\
v(Y_1) & v(Y_2) & v(y_2)
\end{array} \right\},
\end{eqnarray}
where the vertical bar separates Alice's instructions from Bob's,
is then
\begin{eqnarray}
G := \left\{ \begin{array}{r|rr}
1 & 1 & 1 \\
1 & 1 & 1
\end{array} \right\}.
\end{eqnarray}
This strategy only fails whenever Bob is asked (iv) (i.e.,
in~$1/4$ of the rounds). However, there is a quantum strategy that
never fails: the players can win all the rounds if they share
pairs of photons in the state (\ref{benasque}), and give as
answers the results of the corresponding measurements on their
photons.

In a real experiment for testing Bell's theorem, the low
efficiency of detectors opens the possibility that non-detections
could correspond to local instructions such as ``if $X$ is
measured, then the photon will not activate the detector.'' This
allows local realistic theories to simulate the observed results.
To estimate the detection efficiency required to experimentally
discard these theories, let us introduce a modification to the
rules of the game. Let us allow each player to give no answer
whatsoever in a fraction $1-\eta$ of the rounds. If any of the
players does not answer, then that round is not valid. This
modification opens the possibility of the players also using a
fraction of sets of local instructions like
\begin{eqnarray}
B_1 := \left\{ \begin{array}{r|rr}
1 & 1 & 1 \\
0 & 1 & 1
\end{array} \right\}
\end{eqnarray}
or
\begin{eqnarray}
B_2 := \left\{ \begin{array}{r|rr}
1 & 1 & 1 \\
1 & 0 & 1
\end{array} \right\},
\end{eqnarray}
where $0$ means that the corresponding player will not answer the
corresponding question. For instance, if the players share $B_1$,
Alice will not answer question (II); and if they share $B_2$, Bob
will not answer questions (iii) and (iv).

Let us suppose that the players are using sets of predefined
answers (or, equivalently, that the observed data can be described
by a local realistic theory). For instance, sets like $G$ with a
frequency $1-p$, sets like $B_1$ with a frequency $p/2$, and sets
like $B_2$ with a frequency $p/2$, where $p$ depends on the
efficiency of the photodetector corresponding to photon~$j$,
\begin{equation}
\eta_j = 1 - p + \frac{p}{2} f_j + \frac{p}{2},
\label{eta}
\end{equation}
where $f_1$ ($f_2$) is the probability that Alice (Bob) answers
[i.e., she (he) does not get the instruction $0$ in her (his) set]
when they are using a $B_1$ ($B_2$) set. In our case $f_j = 1/2$.

Let us calculate the minimum detection efficiency required to
discard the possibility that the players are using this particular
set of predefined answers (or, equivalently, that the observed
data can be described by a local realistic theory). Then, to
simulate the quantum probability of winning the game, the minimum
value of $p$ is obtained by solving the equation
\begin{equation}
P_Q = (1-p) P_G + \frac{p}{2} P_{B_1} + \frac{p}{2} P_{B_2},
\label{PQ}
\end{equation}
where $P_Q$ is the quantum probability of winning the game, $P_G$
is the probability of winning the game when the players use a $G$
set, and $P_{B_j}$ is the probability of winning when the players
use a $B_j$ set and both answer the questions. In our case,
$P_Q=1$, $P_G=3/4$ and $P_{B_j}=1$. Introducing these values in
Eqs.~(\ref{eta}) and (\ref{PQ}), we arrive at the conclusion that
our local model can simulate the quantum predictions if $\eta_j
\le 3/4 = 0.75$. An exhaustive examination of all possible sets of
local instructions shows that the previously presented model is
indeed optimal and therefore we conclude that local realistic
theories {\em cannot} simulate the quantum predictions if
\begin{equation}
\eta_j > 3/4,
\label{etanew}
\end{equation}
which is the same efficiency needed for a loophole-free experiment
based on the three-observer version of GHZ's
proof~\cite{Larsson98}. Indeed, the efficiency required for a
two-observer AVN proof based on the state~(\ref{benasque}) can be
lowered to $\eta_j > 11/16 \approx 0.69$ if we consider 12 quantum
predictions~\cite{Cabello05}.

Let us compare these efficiencies with that required for a
loophole-free experiment based on the two-photon
version~\cite{CPZBZ03,YZZYZZCP05} of the the two-observer AVN
proof~\cite{Cabello01a,Cabello01b}. To be consistent with
9~specific predictions of quantum mechanics (for details,
see~\cite{Cabello01b}), local realistic theories predict the
following 9 relations between the values of the LERs:
\begin{eqnarray}
v(Z_1) & = & -v(Z_2),
\label{one} \\
v(z_1) & = & -v(z_2),
\label{two} \\
v(X_1) & = & -v(X_2),
\label{three} \\
v(x_1) & = & -v(x_2),
\label{four} \\
v(Z_1 z_1) & = & v(Z_2) v(z_2),
\label{five} \\
v(X_1 x_1) & = & v(X_2) v(x_2),
\label{six} \\
v(Z_1) v(x_1) & = & v(Z_2 x_2),
\label{seven} \\
v(X_1) v(z_1) & = & v(X_2 z_2),
\label{eight} \\
v(Z_1 z_1) v(X_1 x_1) & = & -v(Z_2 x_2) v(X_2 z_2).
\label{nine}
\end{eqnarray}
It is impossible to assign the values $-1$ or $1$ in a way
consistent with all
Eqs.~(\ref{one})--(\ref{nine})~\cite{Cabello01b}. To calculate the
detection efficiency required for a conclusive test based on this
proof we will convert this impossibility into a game in which a
quantum-based strategy beats any classical strategy. Consider
again a team of two players, each of them isolated in a booth.
Alice is asked one out of three possible questions: (I) ``What are
$v(Z_1)$ and $v(x_1)$?,'' (II) ``What are $v(X_1)$ and
$v(z_1)$?,'' or (III) ``What are $v(Z_1 z_1)$ and $v(X_1 x_1)$?''
Analogously, Bob is asked one out of three possible questions: (i)
``What are $v(Z_2)$ and $v(z_2)$?,'' (ii) ``What are $v(X_2)$ and
$v(x_2)$?,'' or (iii) ``What are $v(Z_2 x_2)$ and $v(X_2 z_2)$?''
Each of them must give one of the following answers: ``$-1$ and
$-1$,'' ``$-1$ and $1$,'' ``$1$ and $-1$,'' or ``$1$ and $1$.'' An
interesting feature of this game is that it does not require a
promise: all nine possible combinations of questions are
legitimate. Assuming that all questions are asked with the same
frequency, no classical strategy allows the players to win in more
than~$8/9$ of the rounds. However, the players can win all the
rounds using a quantum strategy~\cite{Cabello01b}. An optimal
classical strategy consists on using sets of instructions
\begin{eqnarray}
\left\{ \begin{array}{cc|cc}
v(Z_1) & v(z_1) & v(Z_2) & v(z_2) \\
v(X_1) & v(x_1) & v(X_2) & v(x_2) \\
v(Z_1 z_1) & v(X_1 x_1) & v(Z_2 x_2) & v(X_2 z_2)
\end{array} \right\}
\end{eqnarray}
like
\begin{eqnarray}
G := \left\{ \begin{array}{rr|rr}
1 & 1 & -1 & -1 \\
1 & 1 & -1 & -1 \\
1 & 1 & 1 & 1
\end{array} \right\},
\end{eqnarray}
which gives correct answers except when Alice is asked
(III) and Bob is asked (iii).

Now let us suppose that each player is also allowed to give no
answer in a fraction $1-\eta$ of the rounds. This is equivalent to
assuming that they can use sets of local instructions like
\begin{eqnarray}
B_1 := \left\{ \begin{array}{rr|rr}
1 & 1 & -1 & -1 \\
1 & 1 & -1 & -1 \\
0 & 0 & 1 & 1
\end{array} \right\},
\end{eqnarray}
or
\begin{eqnarray}
B_2 := \left\{ \begin{array}{rr|rr}
1 & 1 & -1 & -1 \\
1 & 1 & -1 & -1 \\
1 & 1 & 0 & 0
\end{array} \right\}.
\end{eqnarray}
In this case, an optimal ensemble of sets of local instructions
turns out to be $G$ sets with frequency $1-p$, $B_1$ sets with
frequency $p/2$, and $B_2$ sets with frequency $p/2$. Therefore,
$P_Q=1$, $P_G=8/9$, $P_{B_j}=1$, and $f_j=2/3$, which leads us to
conclude that, to avoid the detection loophole in the two-observer
AVN proof in~\cite{Cabello01a,Cabello01b,CPZBZ03}, we need
\begin{equation}
\eta_j > 5/6 \approx 0.83,
\end{equation}
which is higher than the value, given by inequality
(\ref{etanew}), required for the two-observer AVN proof introduced
in this Letter.

AVN proofs follow directly from perfect correlations predicted by
quantum mechanics. However, in a laboratory realization of the
experiment, the observed correlations will not be as perfect as
the proof requires. It is therefore convenient to derive Bell's
inequalities whose validity is necessary for the observed
correlations to be consistent with a very general probabilistic
local realistic theory, which are violated by the quantum
predictions by an amount allowing significant room for the
blurring effect of the imperfections in real experiments. The
relevant features of the AVN proof derive from the fact that, for
the state~(\ref{benasque}),
\begin{equation}
\langle \psi |X_1 X_2 z_2-Y_1 Y_2 z_2+X_1 x_1 Y_2 y_2 +Y_1 x_1 X_2
y_2| \psi \rangle = 4,
\end{equation}
while, as can be easily checked, in any local realistic theory,
this expected value must satisfy
\begin{equation}
|\langle X_1 X_2 z_2-Y_1 Y_2 z_2+X_1 x_1 Y_2 y_2 +Y_1 x_1 X_2 y_2
\rangle| \le 2.
\label{MerminCabello}
\end{equation}
Inequality (\ref{MerminCabello}) resembles Mermin's inequality for
three observers~\cite{Mermin90b}, in the sense that quantum
mechanics predicts a violation of 4, which is indeed the maximum
possible violation of inequality (\ref{MerminCabello}), while the
local realistic limit is 2. However, inequality
(\ref{MerminCabello}) requires only two observers, as in the
original Bell inequality~\cite{Bell64}.

Summing up, we have introduced an AVN proof which combines the
most interesting features appearing, separately, in previous AVN
proofs: it is simple (the contradiction to EPR LERs follows from
only 4 quantum predictions), provides a greater amount of evidence
against local realism (only 3 out of these 4 predictions can be
reproduced by LERs), and requires only two observers. In addition,
we have shown that a conclusive experimental test of this proof
would require lower efficiency detectors than those needed for the
previous two-observer AVN proof, and we have derived the
corresponding Bell inequality for real experiments, which
resembles Mermin's inequality for three observers but requires
only two observers.




The author thanks A.~Broadbent, E.~Galv\~{a}o, A.~Lamas-Linares,
J.-\AA.~Larsson, C.-Y.~Lu, S.~Lu, E.~Santos, and H.~Weinfurter for
useful comments, and acknowledges support by Projects
No.~BFM2002-02815 and No.~FQM-239.



\end{document}